\documentclass[journal]{IEEEtran}

\usepackage{amsmath,amsfonts}
\usepackage{amssymb}
\usepackage{algorithm2e}
\usepackage{array}
\usepackage{textcomp}
\usepackage{stfloats}
\usepackage{url}
\usepackage{verbatim}
\usepackage{graphicx}
\usepackage{cite}
\usepackage{hyperref}
\usepackage{url}
\usepackage{booktabs}
\usepackage{subfigure}
\usepackage{braket} 
\usepackage[dvipsnames]{xcolor}
\usepackage{flushend}
\usepackage{enumerate}
\usepackage{dsfont}

\usepackage{xcolor}
\usepackage{soul}

\SetKwComment{Comment}{/* }{ */}
\RestyleAlgo{ruled}

\usepackage{xcolor} 

\usepackage{multicol}
\usepackage{multirow}

\usepackage{mwe}
\usepackage{float}
\usepackage{balance}
\usepackage{tikz}
\usetikzlibrary{quantikz}
\usepackage{xfrac}
\usepackage{amssymb}

\usepackage{todonotes}

\begin{document}
\title{Enriching Earth Observation labeled data with Quantum Conditioned Diffusion Models  
}
\author{Francesco Mauro$^1$,
\thanks{$^1$Engineering Department, University of Sannio, Benevento, Italy. Email: \{f.mauro@studenti., ullo@\}unisannio.it} Francesca De Falco$^2$, 
\thanks{$^2$Department of Information Engineering, Electronics and Telecommunication, Sapienza University of Rome, Rome, Italy. Email: \{francesca.defalco, andrea.ceschini, massimo.panella\}@uniroma1.it}~Lorenzo Papa$^3$,\thanks{$^3$$\Phi$-lab, European Space Agency, Frascati, Italy. Email: \{Lorenzo.Papa, Alessandro.Sebastianelli\}@esa.int, asebastianelli@ieee.org}~Andrea Ceschini$^2$,~Alessandro Sebastianelli$^{3,4}$\thanks{$^4$ CMCC Foundation - Euro-Mediterranean Center on Climate Change, Italy.},
~\\Paolo Gamba$^5$ 
\thanks{$^5$Department of Electrical, Computer and Biomedical Engineering, University of Pavia, Pavia, Italy. Email: paolo.gamba@unipv.it}, ~Massimo Panella$^2$ and Silvia Ullo$^1$ 
}

\date{}
\maketitle
\markboth{Submitted to IEEE Transactions on Geoscience and Remote Sensing}{}

\begin{abstract}
The rapid adoption of diffusion models (DMs) in the Earth Observation (EO) domain has unlocked new generative capabilities aimed at producing new samples, whose statistical properties closely match real imagery, for tasks such as synthesizing missing data, augmenting scarce labeled datasets, and improving image reconstruction. This is particularly relevant in EO, where labeled data are often costly to obtain and limited in availability.
However, classical DMs still face significant computational limitations, requiring hundreds to thousands of inference steps, as well as difficulties in capturing the intricate spatial and spectral correlations characteristic of EO data. Recent research in Quantum Machine Learning (QML), including initial attempts of Quantum Generative Models, offers a fundamentally different approach to overcome these challenges.
Motivated by these considerations, we introduce the Quanvolutional Conditioned U-Net (QCU-Net), a hybrid quantum--classical architecture that applies quantum operations within a conditioned diffusion framework using a novel quanvolutional feature-extraction approach, for generating synthetic labeled EO imagery.
Extensive experiments on the EuroSAT RGB dataset demonstrate that our QCU-Net achieves superior results. Notably, it reduces the Fréchet Inception Distance by 64\%, lowers the Kernel Inception Distance by 76\%, and yields higher semantic accuracy. Ablation studies further reveal that strategically positioning quantum layers and employing entangling variational circuits enhance model performance and convergence.
This work represents the first successful adaptation of class-conditioned quantum diffusion modeling in the EO domain, paving the way for quantum-enhanced remote sensing imagery synthesis.
\end{abstract}

\begin{IEEEkeywords}
Quantum Computing, Quantum Deep Learning, Quantum Diffusion Models, Remote Sensing, Earth Observation
\end{IEEEkeywords}

\section{Introduction}

\IEEEPARstart{T}{he} use of diffusion models for generating synthetic satellite images and their corresponding labels is becoming an essential tool in Earth Observation (EO), especially considering the high cost and human effort required to produce annotated datasets. 
\vspace{2cm}

While traditional data augmentation techniques, such as geometric transformations and color adjustments have shown limited capability in capturing the underlying data distributions, mainly due to their reliance on fixed mathematical formulations, recent advancements in deep learning have introduced powerful generative approaches. For instance, artificial intelligence methods such as Generative Adversarial Networks (GANs) and Diffusion Models (DMs) \cite{alzahem2023improving, adedeji2022image, ghelichkhani2023generative} manage to synthesize highly realistic EO imagery.

In more detail, these models have been applied to a wide range of remote sensing tasks such as cloud removal \cite{sebastianelli2022plfm}, hyperspectral and multispectral data generation \cite{abuhani2025generative, alibani2024multispectral}, and super-resolution enhancement \cite{filali2024spatiotemporal}. However, challenges remain in computational efficiency, spectral consistency, and domain generalization \cite{liu2024diffusion, tang2024crs}.
Precisely, although GANs provide fast inference, they often suffer from instability and mode collapse. 
On the other hand, DMs can provide more stable and high fidelity training, but require hundreds to thousands of iterations at inference time. 

Following this research path, recent work explored hybrid architectures that combine previous convolutional and vision transformer techniques \nobreak into GANs and DM strategies to balance real\nobreak ism, controllability, and computational cost \cite{ghelichkhani2023generative}. Addition\nobreak ally, in the field of EO research, text-to-image generative models, which leverage semantic features, have enabled conditional image generation \cite{sebaq2024rsdiff}. 
However, generative modeling for EO remains only partially explored, with a limited number of studies addressing its full potential.

In parallel, the research field of Quantum Machine Learning (QML) has gained traction with the development of quantum algorithms for supervised and unsupervised learning tasks \cite{farhi2018classification, Schuld_2017, aimeur:hal-00736948, Benedetti_2019}. 
Moreover, quantum generative models, including Quantum GANs (QGANs) \cite{PhysRevApplied.16.024051} and Quantum Diffusion Models (QDMs) \cite{DeFalco2024QuantumHybrid, de2024quantum}, are offering a fundamentally different approach to data generation by leveraging the expressive power of quantum states and the high dimensionality of Hilbert space \cite{Bravyi_2018, Abbas_2021}.

In light of this, \cite{chang2024latentstylebasedquantumgan} recently proposed a hybrid quantum GAN architecture (LaSt-QGAN) applied to the SAT4 EO dataset, marking one of the first applications of quantum generative modeling to EO. 
Furthermore, \cite{de2024quantum} proposed a Quantum Latent Diffusion Model (QLDM) that leveraged the established idea of classical latent diffusion models with a quantum denoiser circuit. 
However, QDMs performances are hindered by the information compression due to the autoencoder process. 
Specifically, in order to implement a fully quantum denoiser circuit, the images must be compressed to extremely low-dimensional representations, such that they can be encoded via angle encoding on a limited number of qubits. 
Nevertheless, as reported in \cite{de2024quantum}, the results are encouraging and support the feasibility of leveraging quantum computing within generative models in few-shot learning scenarios.

Motivated by these considerations, this work proposes 
the \textbf{Quanvolutional Conditioned U-Net (QCU-Net)} model, a
hybrid quantum-classical architecture which applies quantum
operations to feature extraction stages via a novel quanvolutional
approach within a conditioned diffusion framework.
Precisely,
we leverage a class-conditioning mechanism to generate class-dependent synthetic images.
This approach, when compared with classical data augmentation techniques, enables the creation of more diverse and semantically rich samples, potentially better suited for enhancing the performance of general-purpose downstream EO tasks. 
Moreover, its value becomes particularly evident in scenarios where labeled data are scarce.

Motivated by these considerations, this work addresses several open challenges in EO, including the scarcity of labeled data, the need for high-quality class-conditioned augmentation, and the difficulty of generating samples that preserve the spatial and spectral properties of real satellite imagery.
The main contributions of this work are the following:

\begin{itemize}
    \item EO-driven motivation: we introduce the first class-conditioned quantum diffusion model specifically designed to generate high-quality labeled EO imagery, enabling effective data augmentation in scenarios where labeled data are scarce or costly to acquire.
    
    \item Hybrid quantum–classical architecture for EO data generation: we propose the QCU-Net, a novel quantum-enhanced framework where the U-Net is enriched with (i) quantum layers at the bottleneck, where spatial resolution is minimal, and (ii) an additional quanvolutional layer placed earlier in the encoder;
    
    \item Improved conditioning and downstream utility: we introduce a class-conditioning mechanism and a dedicated evaluation protocol to verify that generated samples belong to the intended land-cover class. Our approach yields semantically richer and more diverse samples, improving downstream tasks such as classification.
    
    \item Comprehensive evaluation and ablation studies: we benchmark multiple quantum circuit ansatz and configurations to improve generation expressiveness and validate the results via ablation studies.
\end{itemize}

Our \textbf{QCU-Net} model achieves substantial improvements over classical diffusion-based models, as well as other quantum architectures. Notably, QCU-Net achieves a 64\% reduction in Fréchet Inception Distance (FID) and a 76\% reduction in Kernel Inception Distance (KID) compared to classical baselines. It also exhibits enhanced semantic accuracy, particularly in difficult classes such as urban–vegetation boundaries, which are typically prone to misclassification due to high inter-class variability. Finally, our model is shown to produce samples that better approximate the true distribution of satellite imagery compared to a classical counterpart. 

Consequently, by introducing quantum generative modeling to EO, this work aims to lay the groundwork for future quantum-enhanced applications in remote sensing.
These include tasks such as multi-sensor simulation and quantum-enhanced EO data processing, as well as the generation of synthetic data to support general-purpose downstream applications in contexts where labeled data is limited or difficult to obtain. To the best of our knowledge, this is the first succes\nobreak sful implementation of a quantum-enhanced, class-conditioned diffusion approach within the EO domain.

The rest of the paper is organized as follows. Section \ref{sec:background} introduces the fundamentals of quantum computing and DMs, offering the theoretical foundation needed to grasp our hybrid quantum-classical generative technique. Section \ref{sec:related_works} discusses the state of the art, reviewing both classical generative models, including GANs and DMs in EO, and recent developments in quantum generative models. 
Section \ref{sec:methodology} describes our proposed model, presenting in detail the QCU-Net architecture designed for EO imagery synthesis, as well as the other quantum model integrated within the diffusion pipeline used for comparison. We also outline the class-conditioning mechanism that guides generation.
Section \ref{sec:implementation} summarizes the experimental setup, including software, training procedures, and evaluation metrics, along with the specifics of the EuroSAT dataset utilized in our experiments. 
Section \ref{sec:experimental_results} presents a comprehensive set of experimental results, highlighting key improvements in image realism, semantic fidelity, and computational efficiency, as well as the insights derived from a detailed ablation study. 
Finally, Section \ref{sec:conclusion} concludes the paper, summarizing the contributions and proposing future research directions for quantum-enhanced EO applications.


\section{Background} \label{sec:background}
This section introduces the fundamental concepts underlying the two main building blocks of our work. 
Precisely, Section \ref{subsec:f_quantum} presents an overview of quantum computing, while Section \ref{subsec:f_dm} focuses on DMs and their relevance to our methodology.

\subsection{Fundamentals of Quantum Computing} \label{subsec:f_quantum}

Quantum computing harnesses the principles of quantum mechanics to process information in fundamentally new ways. Unlike classical systems based on bits, quantum computers use quantum bits (\textit{qubits}) that exploit superposition and entanglement to enable computational paradigms with potentially exponential speedups for specific tasks \cite{nielsen2010quantum}. 
This section outlines the essential building blocks of quantum computation: quantum states, quantum gates and circuits, measurement, and variational algorithms.

\paragraph{Quantum States}
In quantum computing, a qubit lives in a two-dimensional complex Hilbert space spanned by:
\begin{equation}
\ket{0} = \begin{pmatrix} 1 \\ 0 \end{pmatrix}, \quad
\ket{1} = \begin{pmatrix} 0 \\ 1 \end{pmatrix}
\end{equation}

Its general state is a normalized superposition with complex coefficients  $\alpha, \beta \in \mathbb{C}$:
\begin{equation}
    \ket{\psi} = \alpha\ket{0} + \beta\ket{1}, \quad |\alpha|^2 + |\beta|^2 = 1 
\end{equation}

Alternatively, using real parameters:
\begin{equation}
    \ket{\psi} = e^{i\gamma} \left( \cos\left(\frac{\theta}{2}\right)\ket{0} + e^{i\phi} \sin\left(\frac{\theta}{2}\right)\ket{1} \right)
\end{equation}

An $n$-qubit system lies in a $2^n$-dimensional Hilbert space:
\begin{equation}
    \ket{\psi} = \sum_{x \in \{0,1\}^n} \alpha_x \ket{x}
\end{equation}
with $\ket{\psi_1} \otimes \ket{\psi_2} = \ket{\psi_1\psi_2}$ denoting multi-qubit states. A state is \textit{entangled} if it cannot be written as a tensor product of individual qubit states: entanglement is a uniquely quantum phenomenon in which the state of each qubit cannot be described independently of the others, even when spatially separated. 

\paragraph{Quantum Gates and Circuits}

Quantum operations correspond to unitary transformations, i.e. linear operators $U$ satisfying:
\begin{equation}
    U^\dagger U = UU^\dagger = I
\end{equation}
that preserve the norm of quantum states. These transformations are implemented via quantum gates, the basic building blocks of quantum circuits. Unlike classical logic gates, quantum gates are reversible and operate on complex-valued amplitudes.

Single-qubit gates, summarized in Tab. \ref{tab:single_qubit_gates}, act on individual qubits and are represented by $2\times2$ unitary matrices. They manipulate the state by rotating it around the Bloch sphere. For example, the Pauli gates $X$, $Y$, and $Z$ correspond to $\pi$-rotations about the $x$, $y$, and $z$ axes, respectively. The Hadamard gate ($H$) creates superposition by mapping basis states to equal-weight combinations. Rotation gates $R_x$, $R_y$, and $R_z$ perform arbitrary-angle rotations around the respective axes, with $\theta$ specifying the rotation angle.

\begin{table}[h!]
    \centering
    \caption{Example of single-qubit gate operations: $X$, $Y$, $Z$, $H$, and rotations $R_x(\theta)$, $R_y(\theta)$, $R_z(\theta)$}
    \resizebox{\columnwidth}{!}{
    \begin{tabular}{c|c|c}
        \toprule
        Gate & Matrix & Description \\ \midrule
        $X$ & $\begin{bmatrix} 0 & 1 \\ 1 & 0 \end{bmatrix}$ & Bit flip (rotation about $x$ axis) \\
        $Y$ & $\begin{bmatrix} 0 & -i \\ i & 0 \end{bmatrix}$ & Bit and phase flip (rotation about $y$ axis) \\
        $Z$ & $\begin{bmatrix} 1 & 0 \\ 0 & -1 \end{bmatrix}$ & Phase flip (rotation about $z$ axis) \\
        $H$ & $\frac{1}{\sqrt{2}}\begin{bmatrix} 1 & 1 \\ 1 & -1 \end{bmatrix}$ & Creates equal superposition \\
        $R_x(\theta)$ & $\begin{bmatrix} \cos(\frac{\theta}{2}) & -i\sin(\frac{\theta}{2}) \\ -i\sin(\frac{\theta}{2}) & \cos(\frac{\theta}{2}) \end{bmatrix}$ & Rotation around $x$ axis by angle $\theta$ \\
        $R_y(\theta)$ & $\begin{bmatrix} \cos(\frac{\theta}{2}) & -\sin(\frac{\theta}{2}) \\ \sin(\frac{\theta}{2}) & \cos(\frac{\theta}{2}) \end{bmatrix}$ & Rotation around $y$ axis by angle $\theta$ \\
        $R_z(\theta)$ & $\begin{bmatrix} e^{-i\theta/2} & 0 \\ 0 & e^{i\theta/2} \end{bmatrix}$ & Rotation around $z$ axis by angle $\theta$ \\
        \bottomrule
    \end{tabular}}
    \label{tab:single_qubit_gates}
\end{table}

Multi-qubit gates, summarized in Tab. \ref{tab:two_qubit_gates}, enable interactions between qubits and are essential for generating entanglement. The most common one is the controlled-NOT (CNOT) gate, a two-qubit operation where the second (target) qubit is flipped if and only if the first (control) qubit is in state $\ket{1}$:
\bigskip
\begin{equation}
    \text{CNOT} \ket{00} = \ket{00}, \quad \text{CNOT} \ket{10} = \ket{11}
\end{equation}
Another common gate is the SWAP gate, which swaps the states of two qubits:
\begin{equation}
    \text{SWAP} (\ket{\psi_1} \otimes \ket{\psi_2}) = \ket{\psi_2} \otimes \ket{\psi_1}    
\end{equation}

\begin{table}[h!]
    \centering
    \caption{Example of two-qubit gate operations: CNOT and SWAP.}
    \begin{tabular}{c|c|c}
        \toprule
        Gate & Matrix & Description \\ \midrule
        CNOT & $\begin{bmatrix} 1 & 0 & 0 & 0 \\ 0 & 1 & 0 & 0 \\ 0 & 0 & 0 & 1 \\ 0 & 0 & 1 & 0 \end{bmatrix}$ & Flips target if control is $\ket{1}$ \\
        SWAP & $\begin{bmatrix} 1 & 0 & 0 & 0 \\ 0 & 0 & 1 & 0 \\ 0 & 1 & 0 & 0 \\ 0 & 0 & 0 & 1 \end{bmatrix}$ & Swaps qubit states \\
        \bottomrule
    \end{tabular}
    \label{tab:two_qubit_gates}
\end{table}

These gates can be combined in sequence to build a quantum circuit where arbitrary unitary transformations are applied on multi-qubit systems. In fact, a finite set of gates, such as $\{R_x(\theta), R_z(\theta), \text{CNOT}\}$, forms a universal gate set capable of approximating any unitary operation to arbitrary precision.

\paragraph{Quantum Measurement}
Measurement projects a quantum state onto a classical outcome. A qubit $\ket{\psi} = \alpha\ket{0} + \beta\ket{1}$ collapses to $\ket{0}$ with probability $|\alpha|^2$ and $\ket{1}$ with $|\beta|^2$. In the computational ($Z$) basis, the expectation value with respect to the observable $Z$ is:
\begin{equation}
    \langle Z \rangle = \bra{\psi} Z \ket{\psi} = |\alpha|^2 - |\beta|^2
\end{equation}

More generally, for an observable $\hat{O}$, we compute:
\begin{equation}
    \langle \hat{O} \rangle = \bra{\psi} \hat{O} \ket{\psi}
\end{equation}
Due to measurement stochasticity, repeated samples are required to estimate expectation values, especially in variational and simulation tasks.

\paragraph{Variational Quantum Algorithms}
Variational Quantum Algorithms (VQAs) are hybrid methods leveraging NISQ devices. They consist of data encoding, parameterized evolution, and measurement. Classical input $\mathbf{x} \in \mathbb{R}^n$ is encoded via a unitary $U_{\phi}(\mathbf{x})$:

\begin{equation}
    \ket{\phi(\mathbf{x})} = U_{\phi}(\mathbf{x})\ket{0}^{\otimes n}
\end{equation}

Angle encoding is common, using $\mathbf{x}$ to control rotations on each qubit.

A parameterized ansatz $U_W(\boldsymbol{\theta})$ is then applied, typically composed of rotation and entangling gates. The output obtained as expected value with respect to $\hat{O}$ is:
\begin{equation}
    f(\mathbf{x}, \boldsymbol{\theta}) = \bra{\phi(\mathbf{x})} U_W^\dagger(\boldsymbol{\theta}) \hat{O} U_W(\boldsymbol{\theta}) \ket{\phi(\mathbf{x})}
\end{equation}

This value feeds into a classical cost function, optimized iteratively. Gradients are estimated via the parameter-shift rule:
\begin{equation}
    \nabla_{\theta} f(\mathbf{x}, \theta) = \frac{1}{2} \left[f(\mathbf{x}, \theta + \frac{\pi}{2}) - f(\mathbf{x}, \theta - \frac{\pi}{2})\right].
\end{equation}

\subsection{Fundamentals of Diffusion Models}
\label{subsec:f_dm}
In this section we first introduce some fundamental concepts about diffusion models, with a specific focus on the parameters that may impact generation performance.

In particular, diffusion models are characterized by two main phases: the forward (diffusion) process and the reverse process.

During the forward process, the noise schedule \(\beta_t\), as defined in~\cite{sohl2015deep}, governs a Markov chain that can be described as:
\[
(\mathbf{x}_{0:T}) = q(\mathbf{x}_0)\prod_{t=1}^{T}q(\mathbf{x}_{t}|\mathbf{x}_{t-1})\,\]
\[q(\mathbf{x}_{t}|\mathbf{x}_{t-1})= \mathcal{N}\left(\sqrt{1 -\beta_{t}}\mathbf{x}_{t-1}, \beta_{t}\mathbf{I}\right)\
\]
Data samples \(\mathbf{x_0} \sim q\) from the true distribution are gradually corrupted through \(T\) steps, resulting in ${\mathbf{x}_T\sim \mathcal{N}(\mathbf{0}, \mathbf{I})}$.

The reverse process instead has the aim of inverting this diffusion by sampling from ${q(\mathbf{x}_{t-1}|\mathbf{x}_t)}$. However, this conditional distribution is generally intractable due to its dependence on the true data distribution, and thus a data-driven approach is adopted.

Let $p_{\boldsymbol{\theta}}$  be a neural network-based approximation of   ${q(\mathbf{x}_{t-1}|\mathbf{x}_t)}$, modeled as:
\[
p_{\boldsymbol{\theta}}(\mathbf{x}_{t-1}|\mathbf{x}_{t})= \mathcal{N}\left(\boldsymbol{\mu}_{\boldsymbol{\theta}}(\mathbf{x}_{t},t), \boldsymbol{\Sigma}_{\boldsymbol{\theta}}(\mathbf{x}_{t},t)\right)\
\]
In simplified settings, ${\boldsymbol{\Sigma}_{\boldsymbol{\theta}}(\mathbf{x}_{t},t)}$ is kept fixed equal to ${\beta_t\mathbf{I}}$, and the network is trained to predict the noise term $\boldsymbol{\epsilon}_{\boldsymbol{\theta}}(\mathbf{x}_{t},t)$, since the mean $\boldsymbol{\mu}_{\boldsymbol{\theta}}(\mathbf{x}_{t},t)$  can be expressed as a function of the predicted noise as follows: 

\[
\boldsymbol{\mu}_{\boldsymbol{\theta}}(\mathbf{x}_{t},t) = {\frac{1}{\sqrt{\alpha_t}}} \left(\mathbf{x}_t - \frac{1-\alpha_t}{\sqrt{1-\bar\alpha_t}}\boldsymbol{\epsilon}_{\boldsymbol{\theta}}(\mathbf{x}_{t},t)\right)\,.
\]

As a result, the training process is driven by the following loss function:

\[
L^\mathrm{simple}_t = \mathbb{E}_{\mathbf{x}_0\sim q,t,\boldsymbol{\epsilon}\sim\mathcal{N}(\mathbf{0},\mathbf{I})} \left[ \left\| \boldsymbol{\epsilon}_{\boldsymbol{\theta}}(\mathbf{x}_{t},t) - \boldsymbol{\epsilon} \right\|^2 \right]\
\]
with the configuration of the diffusion rate $\beta$ fundamental for its final performance. Different solutions are adopted. In particular, in~\cite{denoising_diff_mod, sohl2015deep} authors set a \textit{linear} $\beta$ variance ranging from $\beta_1 = 10^{-4}$ to $\beta_T=0.02$ with $T=1000$ steps. In contrast, in \cite{nichol2021improved}, authors propose to improve diffusion models with a reparametrization of the generation process variance, i.e., replacing the linear schedule with a squared \textit{cosine} to prevent abrupt changes of noise levels. 
This choice leads to a slower forward process with $T=4000$ steps while increasing reconstructed image details.

\section{Related Works}\label{sec:related_works}

This section reviews the evolution of generative methods for EO image‑synthesis, from early GAN-based approaches (in Section \ref{subsec:sota_gan4eo}) to classical DMs (in Section \ref{subsec:sota_dm4eo}). Additionally, recent quantum generative techniques are also investigated in Section \ref{subsec:sota_qdm}.

\subsection{Generative Adversarial Networks for EO Image Synthesis}\label{subsec:sota_gan4eo}

Generative models, particularly GANs, have become central to EO image synthesis tasks such as cloud removal, super-resolution, hyperspectral generation, and data augmentation. Early augmentation methods based on geometric or radiometric transformations \cite{kamel2021data} lacked the ability to capture the underlying data distribution, limiting variability and realism.

GANs introduced adversarial learning to model complex EO data distributions and have been successfully applied across multiple EO domains. Conditional GANs (cGANs) and CycleGANs have enabled both paired and unpaired image translation, including for tasks like cloud-to-clear image mapping \cite{Sarukkai2020CloudRemoval, ebel2020cloud, sebastianelli2022plfm}, SAR-to-optical conversion \cite{xu2022glf}, and temporal gap filling \cite{gonzalez2025generative}. Attention-based architectures, such as transformer-enhanced GANs, have improved image sharpness and spatial consistency in thin-cloud removal \cite{han2024thin} and super-resolution \cite{liu2025remote}.
More recently, He et al. \cite{he2025novel} proposed an Optimal‑Transport GAN (OT‑GAN) that couples dual attention modules with a semi‑discrete OT loss to mitigate large intra‑class and small inter‑class variance in EO scenes, achieving state‑of‑the‑art fidelity on LoveDA and GID‑15 benchmarks.

Beyond image-to-image translation, GANs have also been adapted for task-specific data augmentation. 
For instance, Xie et al. \cite{xie2023gan} proposed a GAN-based sub-instance augmentation framework 
designed to generate realistic fine-scale changes in open-pit mining environments. Their method leverages 
local GAN-based editing to synthesize plausible bitemporal variations, significantly improving change 
detection performance under data scarcity conditions.

In hyperspectral and multispectral settings, GANs have been employed for super-resolution and pansharpening \cite{wang2024hypergan}, as well as for class-conditioned data generation to augment small or imbalanced datasets \cite{sun2022ac}. These approaches demonstrate GANs' strength in producing visually realistic and geometrically detailed outputs, often improving downstream classification or detection performance.

Recent advances have also explored single-image generative learning to address the scarcity of 
large annotated EO datasets. Morphologic-Structure-Aware GAN (MOGAN) \cite{chen2023mogan} 
introduces a morphology-preserving adversarial framework capable of synthesizing realistic 
variations from a single input image while maintaining structural coherence and semantic 
consistency. This approach demonstrates that even without large training collections, 
structure-aware generative modeling can support EO data enrichment and downstream tasks, 
especially in domains where annotated samples are extremely limited.

However, GAN-based generation in EO still suffers from challenges such as training instability, mode collapse, and spectral inconsistency \cite{adedeji2022image, yuan2025empirical}. Moreover, generated outputs may introduce artifacts features, making interpretability and physical validity critical concerns in operational contexts. Consequently, in order to overcome such limitations, our proposed strategy leverage the use of DMs for a more stable and coherent EO image generation.

\subsection{Diffusion Models in Remote Sensing}\label{subsec:sota_dm4eo}

In the context of EO, DMs have emerged as leading generative approaches, demonstrating a valuable contribution across a wide range of tasks, including image generation \cite{sastry2024geosynth,pang2024hsigene }, super-resolution, cloud removal \cite{tang2024crs, sanguigni2023diffusion}, landscape classification, as well as change and climate prediction \cite{pang2024hsigene}.

In particular, image generation plays a crucial role in EO applications, especially in scenarios involving hyperspectral imaging, where data scarcity is primarily due to the high acquisition cost~\cite{pang2024hsigene}. Similarly, in domains such as digital elevation models, data loss often results from limitations in LiDAR data capture systems~\cite{lo2024diff}.

State-of-the-art generative models like DMs help overcome these limitations, enabling improved performance not only in data reconstruction but also in subsequent downstream tasks~\cite{zhou2024controlcity}.
Recently, \cite{sousa2024data} introduced a four‑stage diffusion‑based augmentation pipeline, combining meta‑prompting, InstructBLIP captioning, and LoRA fine‑tuning of Stable Diffusion within an iterative synthesis loop that boosts CLIP top‑1 accuracy on EuroSAT dataset.
Among the most advanced models in this field, DiffusionSat~\cite{Khanna2024DiffusionSat} and HyperLDM~\cite{liu2023diverse} enhance spectral and spatial consistency through pixel-space and latent-space diffusion architectures, respectively. COP-GEN-Beta ~\cite{Haeberle2025CopGen} extends diffusion modeling to the multimodal EO domain by learning joint distributions across SAR, optical, and elevation modalities.

Nonetheless, training and deploying DMs presents several challenges, primarily due to the need for highly parameterized neural networks, which are computationally intensive and difficult to train \cite{sebaq2024rsdiff}.

Some EO-specific solutions have been proposed in~\cite{cao2024litedit}, such as an efficient diffusion transformer via self-attention distillation \cite{survey_eff_vit}, which significantly reduces the number of parameters. Another approach involves using latent-space diffusion models~\cite{liu2023diverse}, which mitigate computational costs by avoiding pixel-space processing.


\subsection{Quantum Generative Models in Machine Learning}\label{subsec:sota_qdm}

Quantum generative models leverage quantum mechanics to encode and generate complex data distributions. Quantum GANs (QGANs) \cite{farhi2018classification, Benedetti_2019}, Quantum Variational Autoencoders, and Quantum Born Machines \cite{hibat2024framework} were among the first proposed architectures, showing promise in low-data regimes.

Recently, QDMs have been introduced as quantum analogs of DDPMs. The Quantum Generative Diffusion Model (QGDM) \cite{Chen2024QGDM} defines a forward noisy process and a learnable reverse process via quantum circuits. QuDDPM \cite{zhang2024generative} further extends this with parameterized quantum noise injection and denoising for data generation. Hybrid quantum-classical approaches \cite{DeFalco2024QuantumHybrid} combine classical neural networks with quantum layers to enhance expressive power and reduce the number of parameters. Other proposals, such as quantum-noise-driven diffusion \cite{parigi2024quantum}, treat quantum decoherence as a generative feature. 

 Although GANs and diffusion models are increasingly applied to EO data and general-purpose QGANs/QDMs are gaining traction, the application of QDMs to EO tasks remains largely unexplored with, to the best of our knowledge, only one prior work introducing a QDM, specifically a quantum latent diffusion model (QLDM) that couples the classical latent-DM paradigm with a quantum denoiser circuit \cite{de2024quantum}, highlighting a clear opportunity  
 that we addressed by proposing the first class-conditioned QDM for EO image generation.\\
 The novel proposed  QCU-Net model introduces a framework well suited to EO high dimensionality, spectral complexity, and label scarcity data, and with respect to \cite{de2024quantum}, there is no reduction in quality due to compression of images, and moreover, the class-conditioned new model is able to generate labeled images for the different classes. The results offer a powerful tool for the EO domain and community of remote sensing researchers.


\section{Methodology}\label{sec:methodology}
In this section we present the QCU-Net, the quantum–hybrid architecture we propose for EO image synthesis, inspired
by~\cite{DeFalco2024QuantumHybrid,de2024quantum}, as promising direction for the integration of quantum generative models, and more specifically QDMs, into the EO domain.

Moreover, due to the lack of quantum-enhanced comparative baselines, we also implement and describe in this section, another simpler conditioned model 
which we identify as QVCU-Net. 
Therefore, this section is organized as follows:
\begin{enumerate}
    \item \textbf{QVCU-Net} (Section~\ref{sec:qvunet}): it integrates quantum layers only at the bottleneck, providing a point of comparison.
    \item \textbf{QCU-Net} (Section~\ref{sec:quanvunet}): our main model. It enriches the U-Net with (i) quantum layers at the bottleneck, where spatial resolution is minimal, and (ii) an additional quanvolutional layer placed earlier in the encoder \cite{henderson2019quanvolutional,10945995}.
\end{enumerate}

Furthermore, in Section~\ref{ansatz} we provide a more detailed explanation of the quantum variational circuits and the ansatz employed in the two architectures. 
Finally, in Section~\ref{conditioning}, we describe in greater detail the protocol used to assess the quality of class-conditional generation.

\begin{figure*}[t]
    \centering
    \includegraphics[width=\linewidth, trim={0 0 3cm 0}, clip]{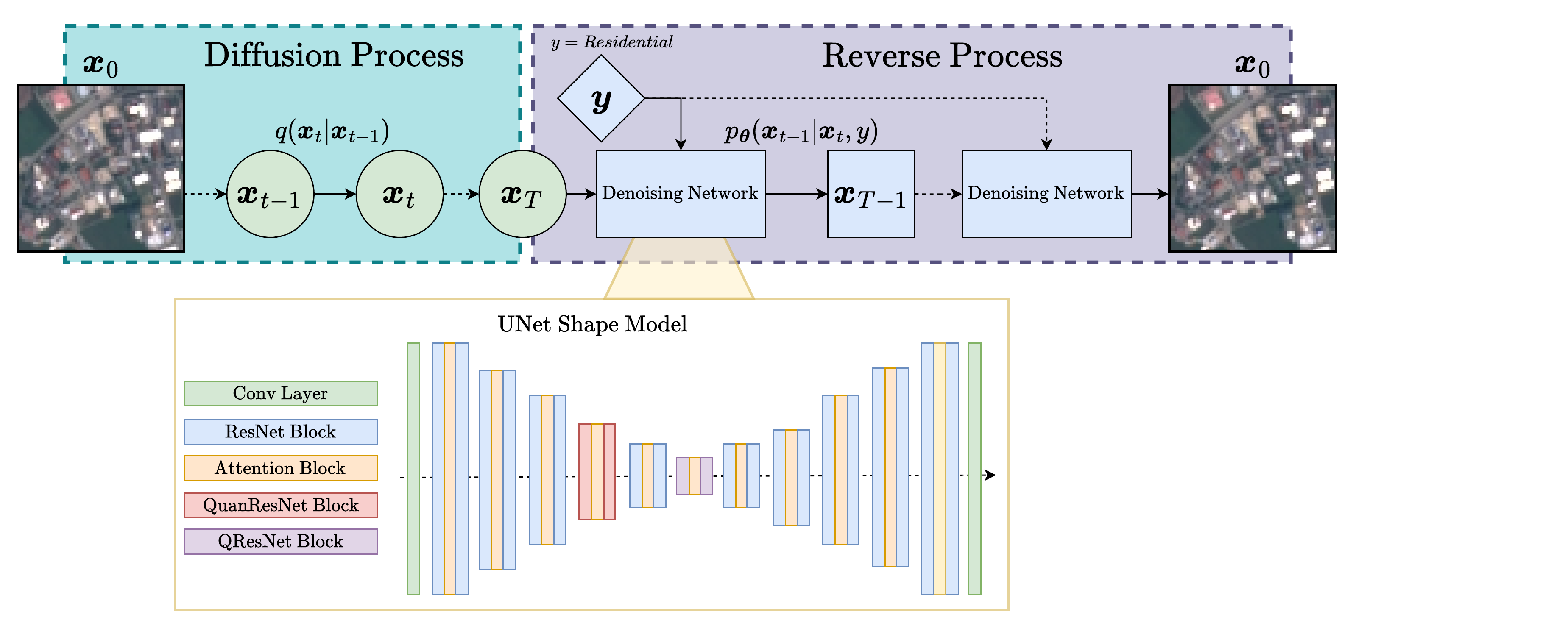}

    \caption{\textbf{Class-conditioned \emph{Quantum Diffusion Model} for EO image generation.}
    \textbf{Top}: the forward diffusion process $q(x_t \mid x_{t-1})$ (in blue) progressively corrupts a clean RGB satellite patch $x_0$ into a Gaussian noise $x_T$. 
    The learned reverse process $p_{\theta}(x_{t-1} \mid x_t, y)$ (in purple) then reconstructs the image step-by-step while being guided by the land-cover label $y$, enabling class-specific generation (e.g., \emph{residential}, \emph{forest}, \emph{river}, \emph{crop}).  
    \textbf{Bottom}: the UNet denoiser integrates both classical components (Convolutional / ResNet / Attention blocks) and two quantum-enhanced residual variants. 
    The \emph{QuanResNet Block} (in red) replaces the first convolution with a quanvolutional quantum filter applied to local spatial patches. 
    While the \emph{QResNet Block} (in purple), placed at the bottleneck of the process, replaces both convolutions with a Variational Quantum Circuit (VQC) acting on $2\times2\times3$ feature blocks and applied to a fraction $\rho$ of channels.
    }
    \label{fig:architecture}
\end{figure*}

\subsection{QVCU-Net}\label{sec:qvunet}

The U-Net, proposed for the first time in \cite{10.1007/978-3-319-24574-4_28} as a neural network for image segmentation, has been established as a standard architecture used for the reverse diffusion process in a DM. Characterized by its distinctive U-shaped structure, it is a highly parameter-rich neural network consisting of an encoder and a symmetric decoder responsible for reconstructing the image from the feature space. 

Starting from a state-of-the-art U-Net architecture, composed of ResNet blocks and Multihead Attention mechanism with four attention heads as suggested in \cite{pmlr-v139-nichol21a}, we integrate quantum layers into this classical architecture, following the approach proposed in \cite{DeFalco2024QuantumHybrid}.  

The rationale behind incorporating quantum layers into a classical network is motivated by the strengths of both paradigms.
Precisely, while classical model can handle high-dimensional data as EO data, which is infeasible for the current NISQ devices, quantum layers offer a potential advantage to overcome some classical issues, e.g. the computational intensity and training difficulties of traditional models. 

Specifically, in the proposed QVCU-Net architecture the quantum component is integrated into the ResNet block located at the vertex as shown in Fig.~\ref{fig:architecture}. 
Given that the image has the lowest spatial dimension at the vertex of the network, few qubits can be used. This motivates the use of angle data embedding. At the same time, having such low spatial dimension allows the implementation of a specific ansatz, described in Section \ref{ansatz}, designed to process groups of adjacent channels (in our case 3). Thereby, the network can exploit both intra-channel and inter-channel correlations to the fullest extent. 

Nevertheless, it is worth noting that the feature representation at the bottleneck consists of 60 channels (given our configuration). 
To better assess the contribution of the quantum layers, we gradually introduced variational quantum circuits over some vertex channels. A detailed ablation study of the impact of the number of quantum layers is provided in Appendix.

Ultimately, compared to the architecture proposed in \cite{DeFalco2024QuantumHybrid}, two main innovations have been introduced: $I)$ the architecture has been adapted to handle EO images, specifically for RGB bands. This adaptation required an increase in the number of blocks within the U-Net, thereby making the model more complex but also more expressive and performative $II)$ conditional generation has been incorporated. In addition to the image and temporal information, the model now also receives as input the class label $y$. This conditioning mechanism is particularly important in the context of the targeted task, i.e. to generate labeled EO images.

\subsection{QCU-Net}\label{sec:quanvunet}
Starting from the QVCU-Net architecture presented in the previous section, 
an additional quantum layer was introduced in the initial part of the encoder, leading to the development of the QCU-Net, illustrated in Fig. \ref{fig:architecture}. 
In fact, the encoder is the portion of the U-Net responsible for extracting hierarchical features and capturing contextual information. Introducing quantum operations in such a critical and sensitive region of the network could therefore significantly enhance overall performance. 

Unlike for the vertex, a key challenge in this encoder branch is that the feature maps have higher dimensions, depending on the ResNet block being used. As a result, encoding the entire feature map into a single variational quantum circuit becomes infeasible, so an alternative approach is required. 

The adopted solution is the quanvolutional approach, inspired by \cite{henderson2019quanvolutional,10945995}. Here, a quantum layer is applied to $k\times k$ patches of the input, akin to a classical convolutional operation, until the entire feature map has been processed, as depicted in Fig. \ref{fig:qlayers}. The key advantage of this method lies not only in the ability to effectively integrate quantum computation into critical stages of the encoder, but also in the construction of a feature map that transforms spatially-local subsections of the input tensor. 

By default the quanvolution operation has been introduced in the $3-rd$ upper layer of the QCU-Net. An ablation study has been conducted and reported in Section \ref{ablation studies} to justify such a placement of the quanvolutional layer within the encoder branch.

\subsection{Variational Quantum Circuits and Ansatz Design}\label{ansatz}

The variational quantum circuits used in both architectures consist of three main components. The first part is dedicated to data encoding. We adopted angle encoding, where classical inputs $\mathbf{x} \in \mathbb{R}^n$ are encoded into quantum states using rotations parameterized by the input values. Specifically, we employ rotations along the $x$-axis, as follows:

\begin{equation}
    \ket{\phi(\mathbf{x})} = R_x(\mathbf{x})\ket{0}^{\otimes n}
\end{equation}

\begin{figure*}[t]
    \centering
    \includegraphics[width=0.98\linewidth]{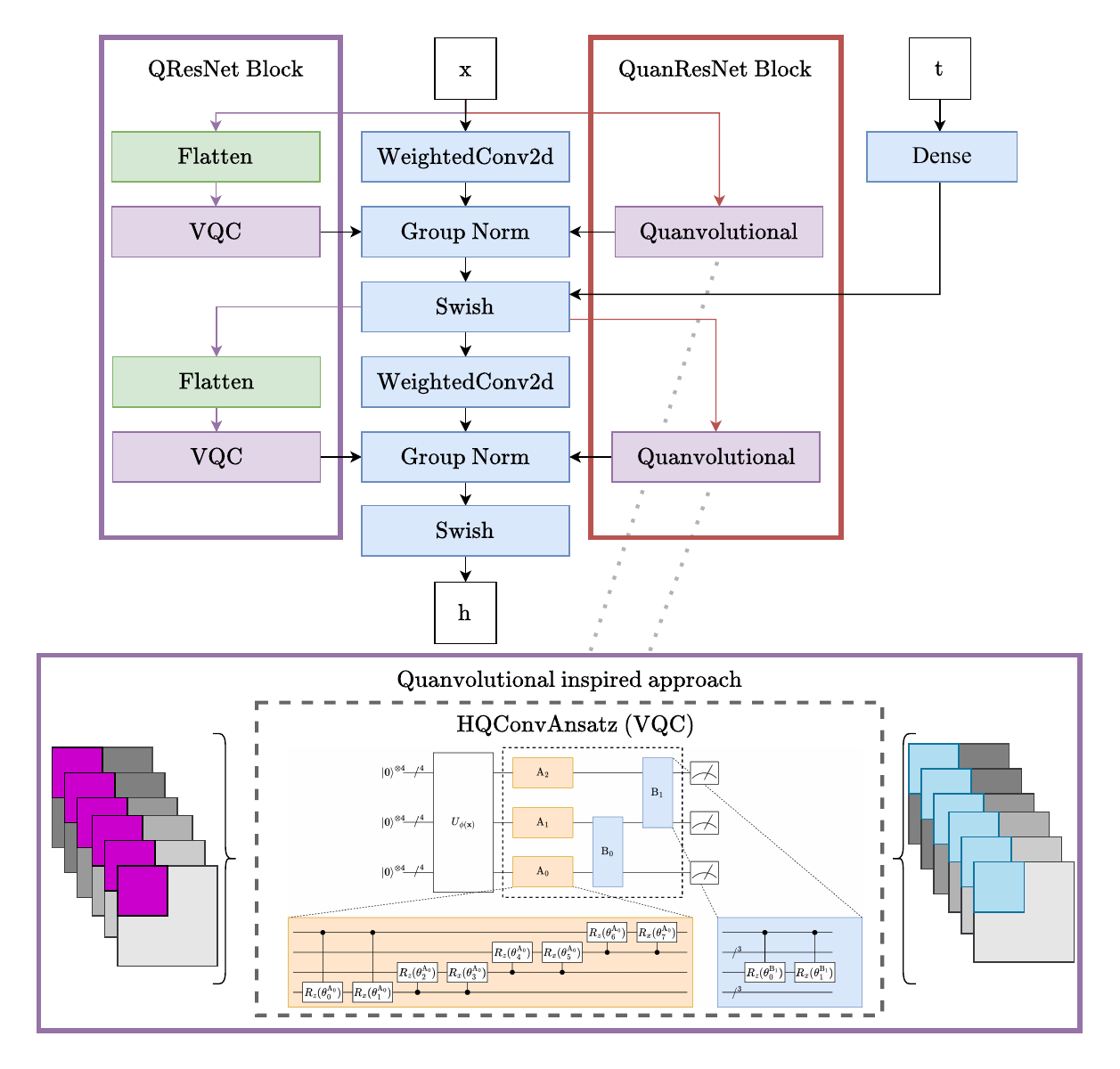}
    \caption{Hybrid quantum--classical ResNet. \emph{Top:} comparison between the classical ResNet block and the QuanResNet block, in which one convolutional layer is replaced by a Quanvolutional (quantum) layer. \emph{Bottom:} schematic of the HQConvAnsatz variational quantum circuit: angle encoding of feature maps, parameterized rotations, entangling gates, and measurements to generate quantum-enhanced features.}
    \label{fig:qlayers}
\end{figure*}

The second part corresponds to the trainable portion of each variational circuit, namely the ansatz. We evaluated three different options: the Hierarchical Quantum Convolutional  (HQConv), the Flat Quantum Convolutional (FQConv), and the Only Rotations ansatz respectively. Fig.~\ref{fig:qlayers} shows the HQConv ansatz, which proves to be the most effective solution, as also explained in Section~\ref{sec:experimental_results}.

The first two ansatz, based on~\cite{Jing_2022}, are particularly well-suited to the application context, where the quantum layer is integrated either at the vertex of a U-Net or within the encoder. These ansatz are specifically designed to process three image channels simultaneously. For example, at the vertex, where the image has spatial dimensions $2 \times 2$, we use a total of 12 qubits ($2 \times 2$ spatial dimensions $\times$ 3 channels). The same principle applies to the QuanvUnet, where the variational quantum circuit acts as a quanvolutional filter, i.e., processing the image in $2 \times 2$ patches and using three channels at a time.

Operating on three channels concurrently allows the model to extract features not only within each individual channel, but also across channels, potentially enhancing feature richness. 

In the following list we better detailed the various ansatzs:

\begin{enumerate}
    \item HQConv (see Fig.~\ref{fig:qlayers}): this ansatz is structured in two stages. The first stage processes intra-channel correlations: each block (labeled Block A in the figure) acts only on qubits belonging to the same channel, and can be expressed as:
    \begin{equation}
        \left|\, q^{o}_{t}, q^o_{t+r}\right\rangle = \left[ CR_x(\theta_{x,t}) \circ CR_z(\theta_{z,t})\right] \left|\, q^{i}_{t}, q^i_{t+r}\right\rangle
    \end{equation}
    
    Where, superscripts $i$ and $o$ indicate the quantum state before and after the application of the block, respectively. The subscript $t$ denotes the control qubit, while $t+r$ indicates the corresponding target qubit.
    The second stage of the ansatz, marked as Block B in Fig.~\ref{fig:qlayers}, is designed to capture inter-channel correlations. It operates as:
    
    \begin{equation}
        \left|\, q^{o}_{t}, q^o_{t+s}\right\rangle = \left[ CR_x(\theta_{x}) \circ CR_z(\theta_{z}) \right] \left|\, q^{i}_{t}, q^i_{t+s} \right\rangle
    \end{equation}
    
    Similarly, superscripts $i$ and $o$ represent the state before and after the application of this block respectively. This structure enables information exchange across different channels, which is particularly advantageous for multi-channel image processing tasks.
    
    \item FQConv: this ansatz is also composed of two stages, but unlike HQConv, it immediately incorporates both intra-channel and inter-channel information at each layer. Both stages use the same general block structure, described by:
    
    \begin{equation}
        \left|\, q^{o}_{t}, q^o_{t+r} \right\rangle = \left[ CR(\theta_{t}) \right] \left|\, q^{i}_{t}, q^i_{t+r} \right\rangle
    \end{equation}
    
    Where, superscripts $i$ and $o$ denote the quantum states before (input) and after (output) the application of the controlled rotation. Precisely, in the first stage, the controlled rotation is performed along the $z$-axis, i.e., $CR_z$, while in the second stage, it is along the $x$-axis, i.e. $CR_x$.

    \item Only-Rotations: this ansatz consists solely of single-qubit rotations along the three axes ($x$, $y$, and $z$), with no entanglement operations. It was included exclusively for comparative analysis purposes, as discussed in Section~\ref{ablation studies}, to evaluate how the more structured ansatz may offer performance improvements, particularly in multi-channel scenarios where entanglement and feature interactions across channels are expected to be beneficial.
\end{enumerate}

The third part of each variational circuit is represented by the measurement stage, where quantum states are mapped back to classical data. In this case, the Pauli-Z observable was used, which has eigenvalues of $+1$ and $-1$.

\subsection{Conditioning Evaluation Protocol} \label{conditioning}

Among the objectives of this work it was necessary to assess whether the proposed QCU-Net and also the implemented QVCU-Net architectures were not only capable of producing visually and quantitatively good samples (for evaluation metrics and results, refer to Sections ~\ref{sec:experimental_results}), but also able to generate images that correctly belong to the desired classes.

To this end, after generating $n$ samples, each associated with a label $y \in \{0, 9\}$, which can be selected at inference time, a state-of-the-art ResNet50 classifier \cite{harini2024resnet,bhatt2024dcrff,bobba2024leveraging,dastour2023comparison} was trained on EuroSAT.

Next, the classifier's weights were frozen, and the $n$ generated samples were fed through the network in inference mode to obtain the predicted labels $\tilde{y} \in {0, 9}$. Assuming the classifier's error rate is minimal (approximately 1\%), the quality of the conditioning was assessed by measuring how often the predicted labels $\tilde{y}$ matched the intended labels $y$ used during sample generation. This evaluation is formalized by the following expression:

\begin{equation}
    \text{Accuracy}_{\text{cond}} = \frac{1}{n} \sum_{i=1}^{n} \mathds{1}_{\{y_i = \tilde{y}_i\}},
\end{equation}

where $\mathds{1}_{\{y_i = \tilde{y}_i\}}$ is the indicator function that returns 1 if $y_i = \tilde{y}_i$, and 0 otherwise.

\section{Experimental Setup}
\label{sec:implementation}
To ensure the reproducibility and clarity of our experiments, this section outlines the complete 
setup used throughout the investigation. In addition, all implementation details, together with 
the pseudocode of the proposed method and the scripts used for training and evaluation, are 
publicly available in our \href{https://github.com/NesyaLab/Quantum-Hybrid-Diffusion-Models-for-EO}{GitHub repository}.
Precisely, in Sections \ref{subsec:software_and_libraries} software environment, libraries,  architectures, and training configurations are respectively reported. 
Subsequently, in Section \ref{subsec:hyperparam}, we present the hyperparameter tuning strategy and provide an overview of the different model sizes evaluated. 
Finally, in Sections \ref{subsec:eval_metrics} and \ref{subsec:dataset_info} evaluation metrics used to assess model performance and the EuroSAT dataset, which serves as the primary benchmark for our experiments, are respectively discussed.

\subsection{Software, libraries and training configurations}
\label{subsec:software_and_libraries}

All code is written in \textit{Python $3.10$}. Core computation relies on \textit{JAX ($0.4.25$)} for accelerated tensor algebra and automatic differentiation, with \textit{Flax ($0.8.2$)} providing the high-level neural-network API. Data ingestion is handled by \textit{TensorFlow Datasets} ($4.9$). Quantum layers are implemented through \textit{PennyLane} ($0.36$) and integrated into custom residual blocks; they are simulated in a noiseless environment due to current computational constraints and are not executed on real quantum hardware. 
The use of \textit{Flax}  in the proposed framework enables a flexible and efficient execution of hybrid neural networks through compilation. In particular, it allows the selection of the \textit{JAX} back-end for the quantum layers implemented in \textit{PennyLane}. This back-end provides access to just-in-time compilation and automatic vectorization of quantum circuits, resulting in a significant reduction in runtime (up to $10 \times$ faster compared to implementations of \textit{PennyLane} within other neural network frameworks such as \textit{PyTorch} ) \cite{chan2024expandinghorizonenablinghybrid}.
Practically, this enables the integration of a significant quantum contribution into a hybrid neural network model without compromising computational efficiency. 

Regarding the evaluation metrics used for image assessment, we employed the \textit{Torchvision} ($0.17$) library \cite{detlefsen2022torchmetrics} for image related metrics and \textit{scikit-learn} ($1.4.1$) \cite{scikit-learn}for classification metrics.
At runtime we load a single \textit{ml\_collections} configuration with batch size 128, learning rate $1{\times}10^{-3}$, cosine $\beta$-scheduler with $T=1000$.
The backbone is a U-Net with self-attention at every resolution; hybrid variants replace selected residual blocks with quantum convolutional modules processing $n=12$ qubits (angle embedding, $L$-layer \textit{FQConv}, \textit{HQConv} or \textit{Only-Rotations} ansatz). Training uses the DDPM cosine $\beta$-schedule with $T=1000$ steps, the $\epsilon$-prediction objective, optional 50\% self-conditioning, and $p^{2}$ loss-weighting. All experiments were executed on the \emph{DEEP Machine} cluster at the Jülich Supercomputing Centre. 
To facilitate replication, we also provide a step-by-step setup guide in the \href{https://github.com/NesyaLab/Quantum-Hybrid-Diffusion-Models-for-EO}{GitHub repository}, including instructions for creating the exact Python environment and running the training 
and evaluation pipelines.

\subsection{Hyperparameters tuning and models' size}
\label{subsec:hyperparam}

\paragraph{Hyperparameter space}

To ensure fair comparisons and establish a strong classical baseline, we performed a hyperparameter tuning of the classical model using the Adam optimizer. The ablation focused on key optimizer parameters: learning rate (lr), first and second momentum coefficients ($\beta_1$, $\beta_2$), and numerical stability term ($\epsilon$). Specifically, we tested the combinations of 3 lr values ($10^{-2}$, $10^{-3}$ and $10^{-4}$), 3 values for $\beta_1$ (1.85, 1.90 and 1.95), 3 values for $\beta_2$ (0.85, 0.90 and 0.95) and 3 values for $\epsilon$ ($10^{-7}$, $10^{-8}$ and $10^{-9}$). Results of the hyperparameters tuning are reported in Section \ref{sec:experimental_results}, paragraph \ref{par:tunig}.

\paragraph{Models' size}

Table \ref{tab:modelsize} illustrates the models' sizes in terms of trainable parameters and the percentage difference compared to the classical model. It is evident that quantum-based models consistently exhibit a reduced number of trainable parameters, with QVCU-Net displaying the fewest. However, our proposed model, QCU-Net, still allows for a reduction in the number of parameters compared to the classical one.

%
 %

\begin{table}[!ht]
    \centering
    \caption{List of models's size in terms of trainable parameters  and the delta in percentage with respect to the classical model. }\label{tab:modelsize}
    \begin{tabular}{lcc}
         \toprule
            Model       & Trainable Parameters & $\Delta_\mathrm{classical}$\%\\
        \midrule
            Classical    & 1.280.843 & -\\
            QVCU-Net & 1.157.723 & 9.6 \\ 
            QCU-Net      & 1.260.939 & 1.6 \\
        \midrule
    \end{tabular}
\end{table}

\subsection{Evaluation Metrics}
\label{subsec:eval_metrics}
The quality of the generated images was assessed using three metrics: \textit{FID}, \textit{KID}, and \textit{Inception Score (IS)}, which collectively evaluate the quality and diversity of generated images \cite{heusel2018ganstrainedtimescaleupdate}.

The FID quantifies the similarity between feature distributions extracted from real and generated images using activations of a pre-trained Inception network. Mathematically, the FID is expressed as:
\begin{equation}
\text{FID}(\mathbf{X}, \mathbf{Y}) = \|\mu_{X} - \mu_{Y}\|^{2}_2 + \text{Tr}\left(\Sigma_{X} + \Sigma_{Y} - 2(\Sigma_{X}\Sigma_{Y})^{1/2}\right),
\end{equation}
where \(\mu_{X}\), \(\mu_{Y}\) and \(\Sigma_{X}\), \(\Sigma_{Y}\) represent the mean and covariance matrices of the feature distributions for real (\(\mathbf{X}\)) and generated (\(\mathbf{Y}\)) images respectively.

The KID evaluates distribution similarity using the squared Maximum Mean Discrepancy (MMD) between features from real and generated images, employing a polynomial kernel defined as:

\begin{equation}
    \centering
    \begin{split}
    &\text{KID}(\mathbf{X}, \mathbf{Y}) =\\ &\mathbb{E}_{x, x' \sim \mathbf{X}}[k(x,x')] + \mathbb{E}_{y,y' \sim \mathbf{Y}}[k(y,y')] - 2\mathbb{E}_{x \sim \mathbf{X}, y \sim \mathbf{Y}}[k(x,y)]
    \end{split}
\end{equation}

with kernel function \(k(x,y) = \left(\frac{1}{d}x^\top y + 1\right)^3\), where \(d\) is the feature dimensionality.

The IS metric quantifies both quality and diversity of the generated images by evaluating their class distribution predicted by the Inception network:
\begin{equation}
\text{IS} = \exp\left(\mathbb{E}_{x \sim \mathbf{Y}}[D_{KL}(p(y|x)\|p(y))]\right),
\end{equation}
where \(D_{KL}\) is the Kullback–Leibler divergence, \(p(y|x)\) the conditional class distribution given an image \(x\), and \(p(y)\) the marginal distribution over generated images.

\begin{figure*}[!ht]
    \centering
    \includegraphics[width=1\textwidth]{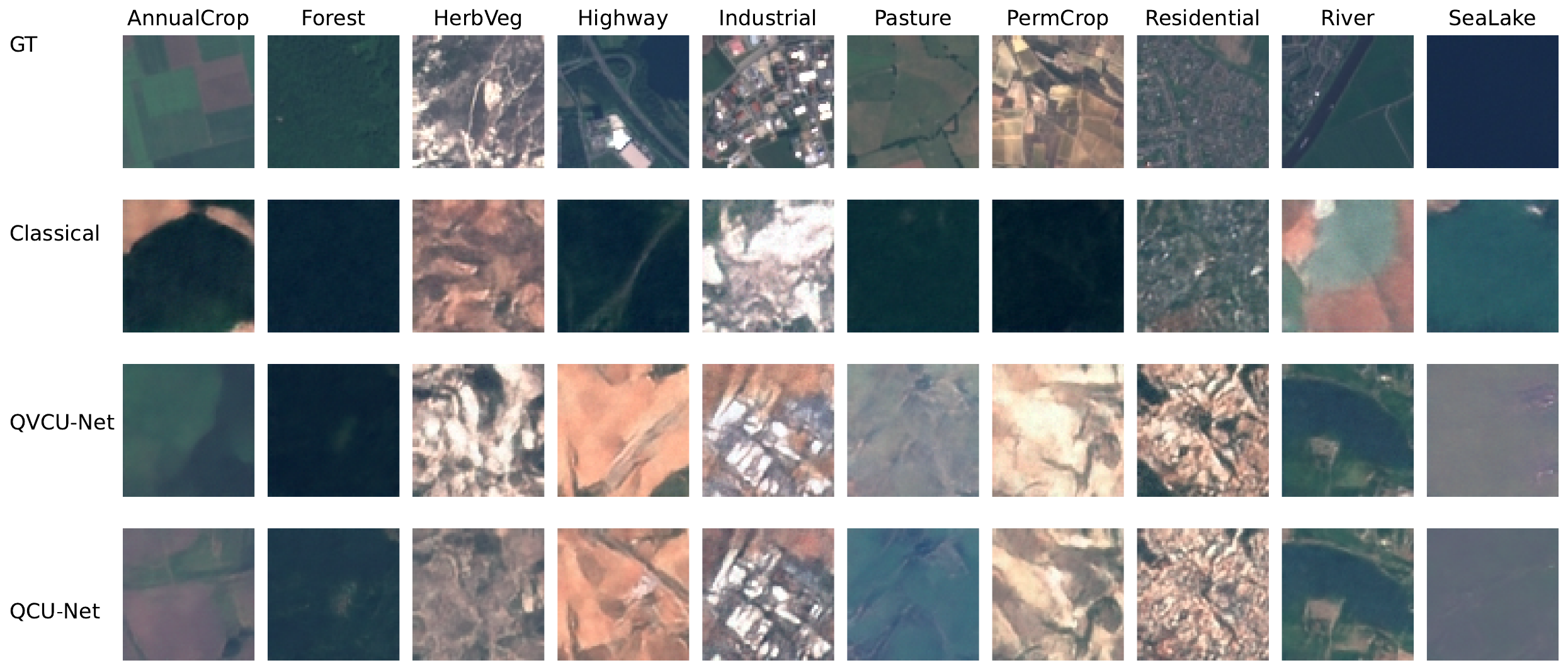}
    \caption{Qualitative comparison between ground truth and generated images on the EuroSAT dataset. Columns correspond (left to right) to the ten land-cover classes: Annual Crop, Forest, Herbaceous Vegetation, Highway, Industrial, Pasture, Permanent Crop, Residential, River, and Sea Lake. Rows show ground truth reference images (first row), outputs of the classical model (second row), and outputs of the proposed quantum model (QCU-Net) (third row).
    }
    \label{fig:qualitative_results}
\end{figure*}

Lower FID and KID values indicate closer alignment with real image distributions, whereas a higher IS value denotes superior quality and greater diversity in the generated samples. In our experimental setup, the metrics were computed using the TorchMetrics library \cite{detlefsen2022torchmetrics}. Specifically, the KID was computed using subsets of 100 image pairs (100 real and 100 generated samples) to estimate the mean and variance, whereas the IS was averaged over 10 equal splits of the generated‑image set. 
Finally, the class‑conditioning capability was assessed by generating scikit‑learn’s classification report, which calculates per‑class precision, recall, F1‑score, and support, along with overall accuracy and the macro and weighted averages of those metrics \cite{SOKOLOVA2009427}; in addition, accuracy was computed separately for each individual class.

\subsection{EuroSAT dataset}
\label{subsec:dataset_info}
The EuroSAT dataset \cite{helber2019eurosat} consists of Sentinel-2 satellite images covering 10 different land cover classes, including agricultural, residential, water bodies, among others. It includes 27,000 labeled images, each with a resolution of $64 \times 64$ pixels and 13 spectral bands. While originally designed for land use and land cover classification, we employ this dataset to generate synthetic images using a quantum diffusion model for data augmentation. Precisely, although the dataset contains 13 spectral bands, we restrict our experiments to the RGB channels, as the evaluation metrics used to assess the effectiveness of the method operate on three-channel images and also for implementation reasons. This way, we adopt RGB images as a controlled benchmark to evaluate model behavior in a simplified setting.

\section{Results} \label{sec:experimental_results}
The following section provides a comprehensive evaluation of our hybrid quantum–classical framework on the EuroSAT benchmark.
The section is organized as follows: Section\ref{sec:main_results} reports the main results by comparing the baseline classical convolutional diffusion model and the QVCU-Net with the proposed QCU-Net, highlighting gains in sample realism, diversity, and label fidelity. 
Specifically, we evaluated three architectures: $I)$ classical diffusion model, a purely classical U‑Net architecture employing ResNet blocks and multi‑head attention, serving as a baseline; $II)$ the QVCU-Net and $III)$ the proposed QCU-Net model.
The proposed model uses a single HQConv circuit. Because the feature maps in this branch are too large to encode in a single variational quantum circuit, a “quanvolutional” approach is adopted: the quantum layer is applied iteratively to $k\times k$ patches of the input—analogous to a classical convolution—until the entire feature map is processed. This scheme enables quantum computation to enrich the hierarchical feature extraction by transforming spatially localized subsections of the tensor. 


Finally, Section~\ref{ablation studies} presents a comprehensive ablation study analyzing the impact of quantum layer placement and circuit ansatz. Building upon these results, Section~\ref{par:additional_studies} offers a deeper investigation into specific factors influencing model behavior, including convergence dynamics and class-wise conditioning accuracy.

\subsection{Main results}\label{sec:main_results}

\textbf{Model Generation Capabilities:}
Table~\ref{tab:classic_vs_quantum_summary} compares the classical diffusion model and the QVCU-Net with the proposed QCU-Net model, using a single HQConv circuit, after $60{,}000$ training iterations on the EuroSAT dataset.

The hybrid solution achieves substantial improvements across key metrics: FID drops from $7.22$ to $2.57$, while KID from $3.4\times 10^{-3}$ to $0.8\times 10^{-3}$, representing relative reductions of $64\%$ and $76\%$, respectively. These improvements indicate that QCU-Net model produces samples that are numerically closer to real images. The improvement of QCU-Net is significant even relative to the QVCU-Net model used as the comparison framework.

\begin{table}[!ht]
    \centering
    \caption{Comparisons of best classical and quantum models after 60,000 iterations on the EuroSAT dataset}
    \label{tab:classic_vs_quantum_summary}
    \begin{tabular}{lccc}
        \toprule
        \textbf{Model} & \textbf{FID} & \textbf{KID} & \textbf{IS} \\
        \midrule
        Classical & 7.22 & $0.0034 \pm 0.0008$ & 1.8837 $\pm 0.0306$ \\
        QCU-Net   & \textbf{2.57} & \textbf{0.0008} $\pm$ 0.0003 & $1.7678 \pm 0.0241$ \\
        QVCU-Net  & 7.60 & $0.0037 \pm 0.0008$ & \textbf{1.9011} $\pm 0.0255$ \\
        \bottomrule
    \end{tabular}
\end{table}

Moreover, we extend the comparison to the few quantum-generative baselines currently applicable to our task, latent diffusion and latent GAN models, since, to the best of our knowledge, no pixel-space quantum diffusion model for EO imagery has been proposed so far. As shown in Table~\ref{tab:tabella_comparison_latent}, QCU-Net markedly outperforms all latent models in FID, KID, and IS, despite its larger parameter count. 

\begin{table*}[!h]
    \centering
    \caption{Comparison Between the Proposed QCU-Net and State-of-the-Art Latent Diffusion Models.}\label{tab:tabella_comparison_latent}
    \begin{tabular}{lcccc}
        \toprule
        Model & Trainable parameters & FID & KID & IS \\
        \midrule
        Classical LDM \cite{de2024quantum} & 330 & 35.28 & $0.0368 \pm 0.0021$ & $1.2659 \pm 0.0053$ \\
        QLDM \cite{de2025quantum} & 342 & 27.69 & 0.0258 $\pm$ 0.0018 & 1.3152 $\pm$ 0.0091\\
        Classical GAN \cite{de2024quantum} & 341 & 33.25 & $0.0311 \pm 0.0012$ & 1.1847 $\pm$ 0.01224 \\
        QGAN \cite{chang2024latentstylebasedquantumgan} & 351 & 33.54 & $0.0384 \pm 0.0013$ & 1.1950 $\pm$ 0.0071 \\
        QCU-Net & 1.260.939  & \textbf{2.57} & \textbf{0.0008} $\pm$ \textbf{0.0003} & \textbf{1.7678} $\pm$ \textbf{0.0241} \\
        \bottomrule
    \end{tabular}
\end{table*}

Visual inspection supports these findings: as shown in Fig.~\ref{fig:qualitative_results}, QCU-Net model generates scenes that are noticeably clearer and more visually aligned with ground-truth samples. In contrast, outputs from the classical baseline tend to suffer from blurriness, reduced spatial coherence, and recurrent artifacts.

\begin{table*}[!ht]
    \centering
    \caption{Comparison of performance metrics for each class across Classical Model, QVCU-Net, and QCU-Net. Best values per row are highlighted in bold.} \label{tab:unified_classification_report}
    \renewcommand{\arraystretch}{1.2}
    \begin{tabular}{lccccccccccccc}
        \hline
        \multirow{2}{*}{Class} & \multicolumn{4}{c}{Classical} & \multicolumn{4}{c}{QVCU-Net} & \multicolumn{4}{c}{QCU-Net} & Support\\
        \cline{2-13}
        & Prec. & Rec. & F1 & Acc. & Prec. & Rec. & F1 & Acc. & Prec. & Rec. & F1 & Acc. & \\
        \hline
        AnnualCrop            & 0.61 & 0.83 & 0.71 & 0.83 & 0.66 & 0.84 & 0.74 & 0.83 & \textbf{0.75} & \textbf{0.92} & \textbf{0.83} & \textbf{0.92} & 1027\\
        Forest                & 0.27 & 0.50 & 0.35 & 0.50 & 0.28 & 0.54 & 0.37 & 0.53 & \textbf{0.88} & \textbf{0.56} & \textbf{0.68} & \textbf{0.55} & 1027 \\
        HerbaceousVegetation & 0.42 & 0.62 & 0.50 & 0.61 & 0.49 & 0.59 & 0.53 & 0.58 & \textbf{0.67} & \textbf{0.90} & \textbf{0.77} & \textbf{0.90} & 1027\\
        Highway               & 0.81 & 0.59 & 0.68 & 0.58 & 0.85 & 0.62 & 0.72 & 0.61 & \textbf{0.91} & \textbf{0.74} & \textbf{0.81} & \textbf{0.73} & 1027\\
        Industrial            & 0.80 & 0.84 & 0.82 & 0.83 & 0.79 & 0.87 & 0.83 & 0.87 & \textbf{0.92} & \textbf{0.93} & \textbf{0.93} & \textbf{0.93} & 1027 \\
        Pasture               & 0.72 & 0.34 & 0.46 & 0.33 & 0.72 & 0.36 & 0.48 & 0.35 & \textbf{0.81} & \textbf{0.83} & \textbf{0.82} & \textbf{0.83} & 1027 \\
        PermanentCrop         & 0.74 & 0.29 & 0.41 & 0.28 & 0.68 & 0.37 & 0.48 & 0.36 & \textbf{0.85} & \textbf{0.66} & \textbf{0.74} & \textbf{0.65} & 1027 \\
        Residential           & \textbf{0.98} & 0.53 & 0.69 & 0.53 & 0.96 & 0.63 & 0.76 & 0.63 & \textbf{0.98} & \textbf{0.97} & \textbf{0.97} & \textbf{0.97} & 1027 \\
        River                 & 0.88 & 0.41 & 0.56 & 0.40 & 0.86 & 0.44 & 0.58 & 0.43 & \textbf{0.90} & \textbf{0.68} & \textbf{0.77} & \textbf{0.68} & 1027 \\
        SeaLake               & 0.58 & 0.98 & 0.73 & 0.98 & 0.61 & \textbf{0.99} & 0.75 & 0.98 & \textbf{0.69} & \textbf{0.99} & \textbf{0.81} & \textbf{0.99} & 948 \\
        \hline
        \textbf{Macro Avg}    & 0.68 & 0.59 & 0.59 & 0.59 & 0.69 & 0.62 & 0.62 & 0.64 & \textbf{0.84} & \textbf{0.82} & \textbf{0.81} & \textbf{0.83} & 948 \\
        \hline
    \end{tabular}
\end{table*}

\begin{figure}[!ht]
    \centering
    \includegraphics[width=1\columnwidth]{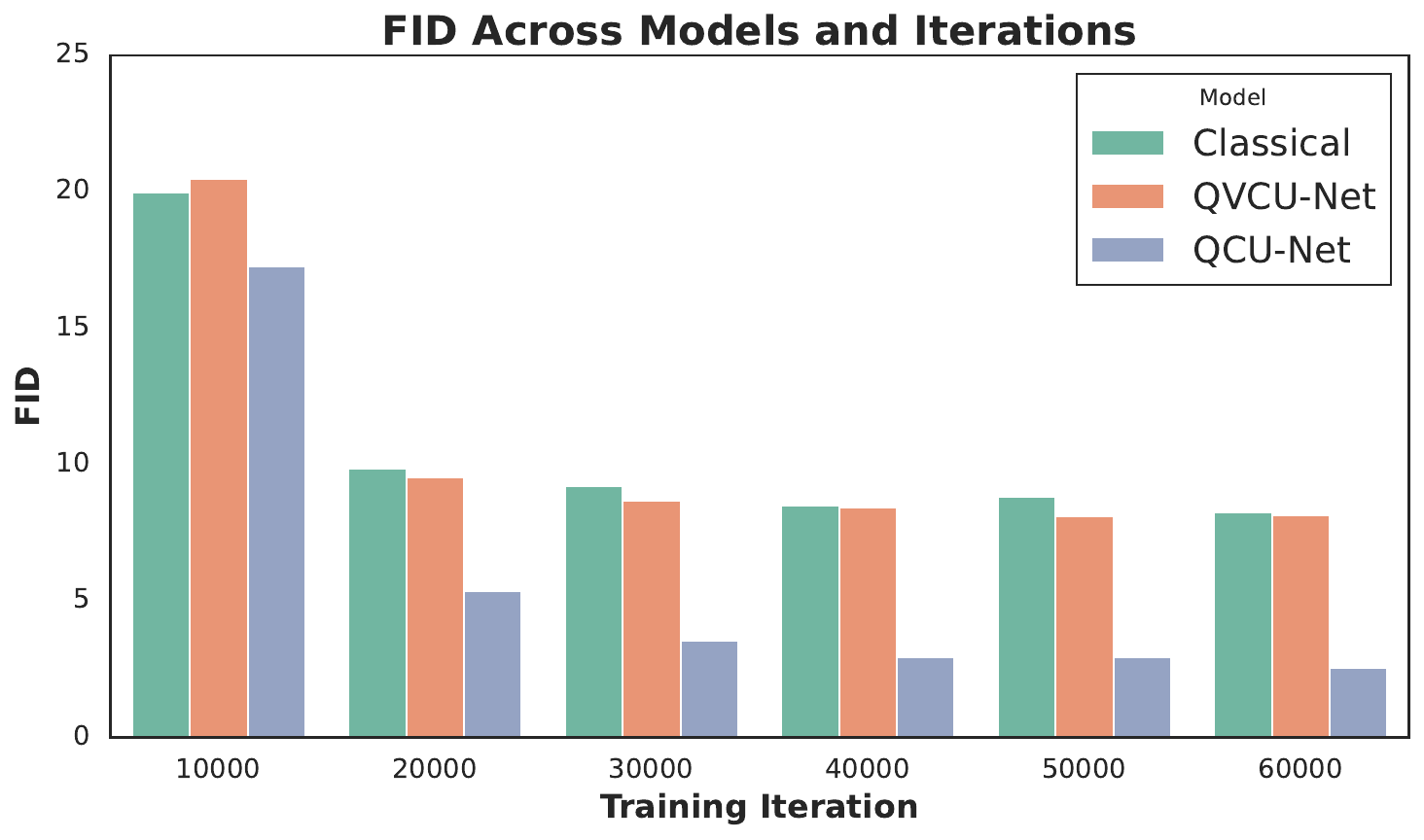}
    \caption{Convergence trends over training iterations for different models and evaluation metrics (FID, KID, IS, and Accuracy). Each plot shows the evolution of performance across iterations for the classical baseline, the QVCU-Net and the proposed QCU-Net. In addition to FID (shown in the main text), similar trends are evident for the other metrics. Notably, qCU-Net not only achieves superior scores but also demonstrates a significantly faster convergence compared to the other approaches.}
    \label{fig:iteration_res}
\end{figure}

Figure \ref{fig:iteration_res} provides a comparative analysis of the FID convergence trends between the classical, QCU-Net and QVCU-Net models, throughout 60.000 optimisation steps. The figure shows a distinct trend: the QCU-Net rapidly surpass the classical baseline and the QVCU-Net. By 20.000 iterations, QCU-Net achieves an FID close to 5, while the classical network remains near 10, and QVCU-Net positions just below 9.5. So it is evident that with 30\% of total iterations the proposed model reaches already a satisfactory result.

QVCU-Net, improves more slowly 
and, as the ablation study will show, the position at which the quantum module is inserted plays a pivotal role in the quality of the generated results.

\textbf{Conditioning Fidelity:}
To evaluate how accurately each model follows the specified class labels, we generate approximately 10.000 synthetic images (1000 per class) and classify them using a ResNet-50 model pre-trained on real EuroSAT data, which achieves a validation accuracy of 98.4\%.

Table \ref{tab:unified_classification_report} shows a clear trend: overall QCU-Net performs best in the class-conditioning task. On average precision it surpasses the classical baseline of 25\% on the precision and 40\% on the recall. Consequently the F1-score follows the same pattern. Conversely, the QVCU-Net offers only a modest step beyond the classical baseline.

\begin{figure*}[t]
  \centering
  \includegraphics[width=.8\textwidth]{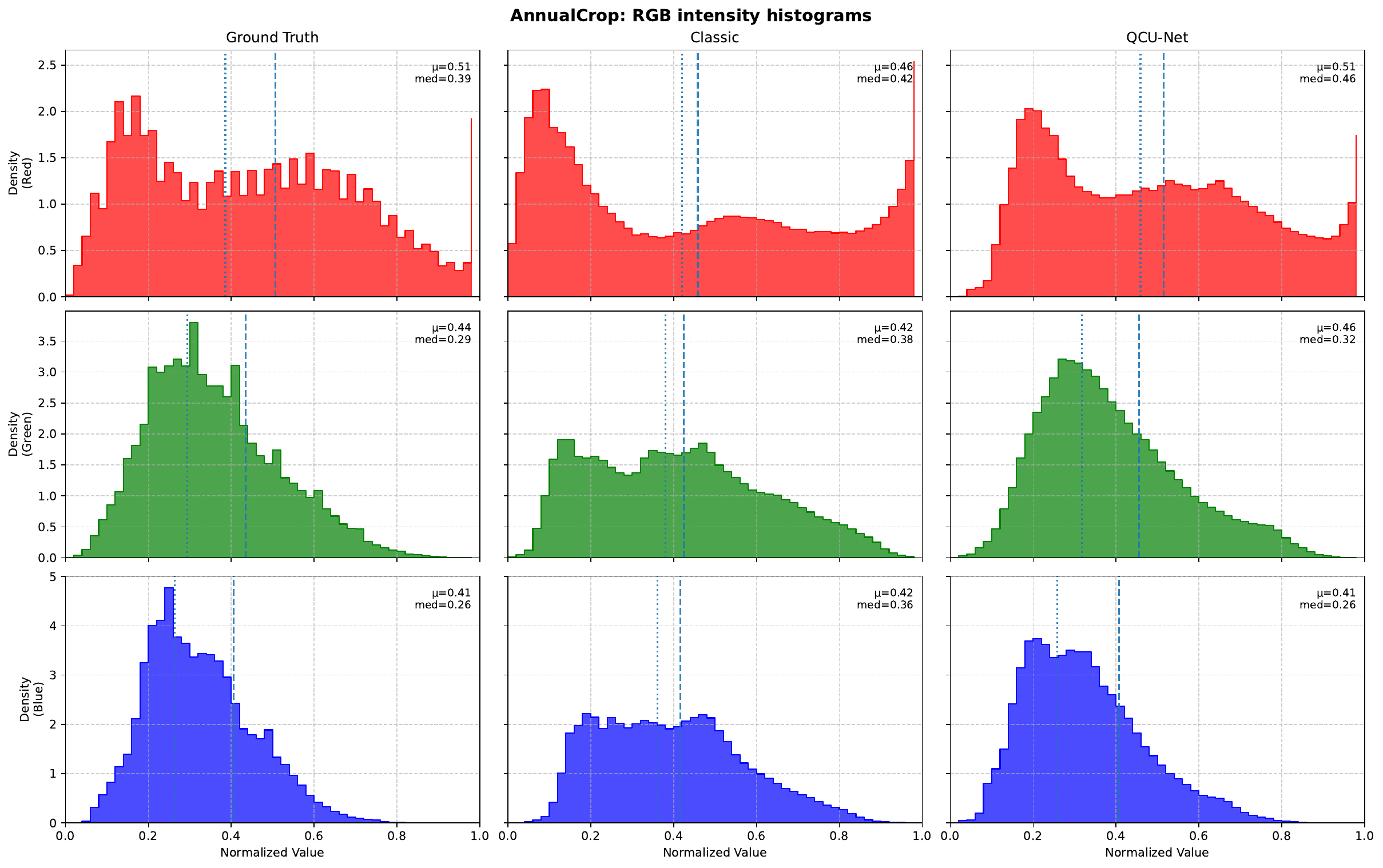}\\[0.6em]
  \includegraphics[width=.8\textwidth]{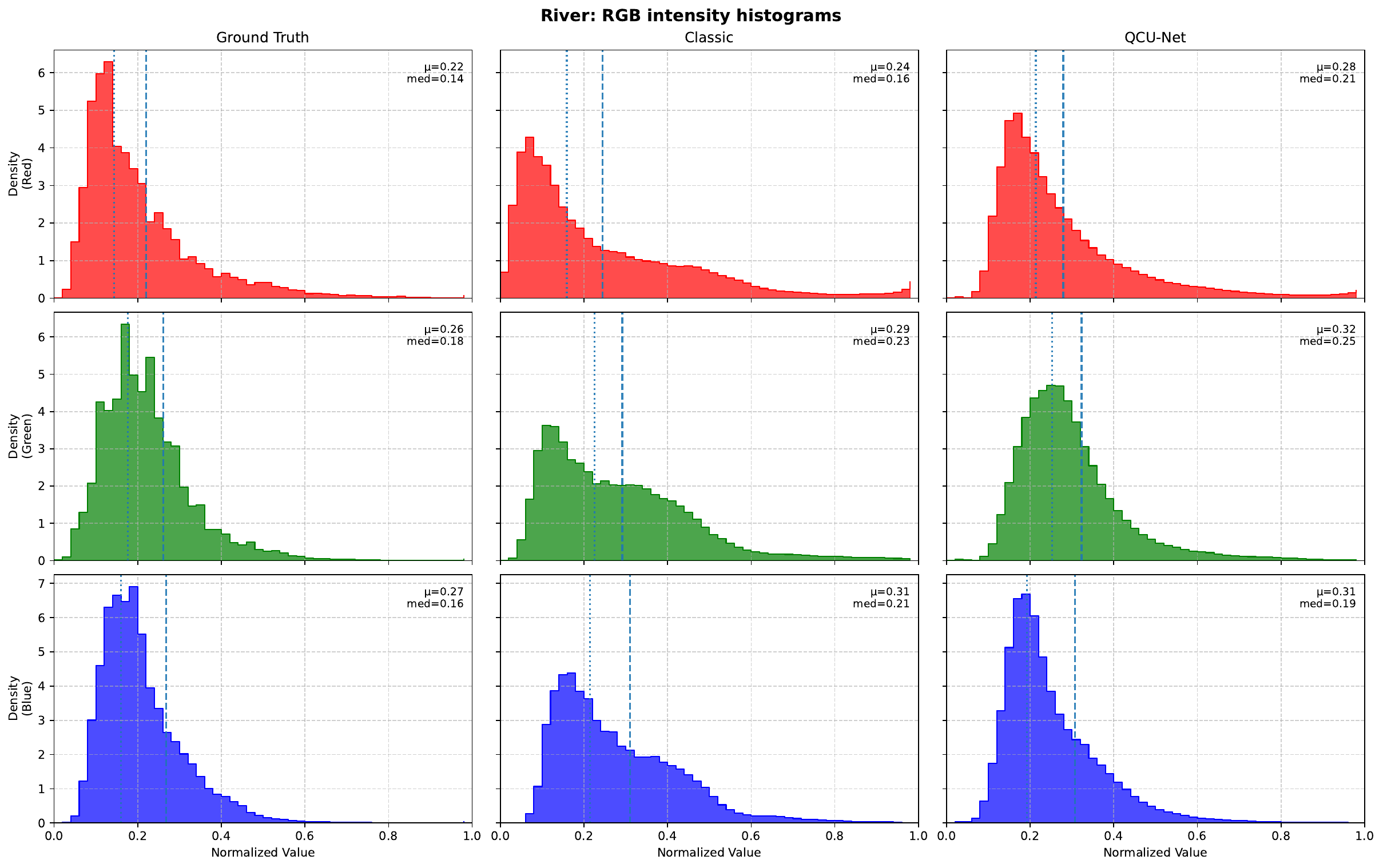}
  \caption{Per-channel intensity distributions (densities; common bins in $[0,1]$) for \textit{AnnualCrop} (top) and \textit{River} (bottom). Vertical dashed and dotted lines denote mean and median. We compare the distribution for the GT, Classic model and the QCU-Net model.}
  \label{fig:hist}
\end{figure*}

\textbf{Preservation of radiometric statistics:}
To evaluate the preservation of the input distribution, we conducted a comparative analysis between the distribution of the GT images and the generated ones, by using a common binning in $[0,1]$, as illustrated in Fig. \ref{fig:hist}. Regarding EO image synthesis tasks, preserving the distribution is crucial because it ensures that the generated images accurately reflect the statistical properties of real-world data. This fidelity is vital for maintaining the integrity and usefulness of the models in practical applications. 
QCU-Net demonstrates greater fidelity in producing an output whose distribution more closely aligns with the original distribution.
Specifically, on \textit{AnnualCrop}, QCU-Net closely follows GT in the blue channel and is consistently closer than the classical diffusion baseline in green; in red, all methods show an accumulation near 1.0. 
Nevertheless, QCU-Net retains a summary statistics (mean and median) nearer to GT than the baseline.
On River, the classical baseline shows a systematic right–shift across the RGB channels (higher mean and median), resulting in over–brightness compared to GT. By contrast, QCU-Net substantially reduces this bias: its red, green, and blue distributions overlap more closely with the GT. Overall, QCU-Net narrows the pixel–level distribution gap.

\subsection{Ablation Study} \label{ablation studies}

\textbf{Position Analysis:}
Table~\ref{tab:position_ablation} presents the results of an ablation study evaluating the impact of quantum layer placement within the QCU-Net architecture. Two configurations were tested: $p_1$, where the quantum layer is applied early in the encoder (via a quanvolutional operation) i.e. $3-rd$ layer, and $p_2$, where it is placed deeper in the network i.e. $4-th$ layer.

Based on the results it can be noticed that the $p_1$ setup yields the best performance across all key metrics: the model achieves the lowest FID = 2.57, the lowest KID = $0.0008 \pm 0.0003$, IS = $1.7678 \pm 0.0241$, and the best classification accuracy on generated samples (81.74\%).

These findings suggest that quantum operations are most effective when applied during the early stages of feature extraction. Therefore, we can consider that quantum feature maps have richer spatial and contextual information. In contrast, the applications of quantum operations to a deeper level of model, results in a decreased performance across all metrics. This suggests that applying quantum layers at this stage may limit the ability of quantum features to contribute meaningfully to representation learning.

\begin{table}[!ht]
    \centering
    \caption{Ablation study on position analysis for QCU-Net after 60{,}000 iterations on the EuroSAT dataset}
    \label{tab:position_ablation}
    \resizebox{\columnwidth}{!}{
    \begin{tabular}{lcccc}
        \toprule
        \textbf{Position} & \textbf{FID} & \textbf{KID} & \textbf{IS} & \textbf{Accuracy (\%)} \\
        \midrule
        $p_1$     & \textbf{2.57} & \textbf{0.0008} $\pm 0.0003$ & \textbf{1.7678} $\pm 0.0241$ & \textbf{81.74} \\
        $p_2$     & 3.32          & $0.0016 \pm 0.0004$ & $1.7400 \pm 0.0242$ & 80.86 \\
        \bottomrule
    \end{tabular}
    }
\end{table}

\textbf{Ansatz Analysis:}
This ablation study focuses on comparing different quantum circuit designs to assess their impact on the quality and fidelity of generated output. Table~\ref{ab:ansatz_comparison} shows results for ansatz presented in Section \ref{ansatz}. Altough, all the tested configuration provide comparable results. best performance are reached with HQConv, producing the lowest FID (2.57), the best KID ($0.0008 \pm 0.0003$) and IS ($1.7780 \pm 0.0241)$.

These results suggest that the HQConv ansatz can be better suited in capturing the high-dimensional dependencies characteristic of EO imagery, likely due to entanglement operations and structured gate composition. In contrast, the Only-Rotations achieves weaker generative metrics, although still performs adequately in terms of Accuracy.

\begin{table}[!ht]
    \centering
    \caption{Ablation study on ansatz choice for QCU-Net after 60{,}000 iterations on the EuroSAT dataset}
    \label{ab:ansatz_comparison}
    \resizebox{\columnwidth}{!}{
    \begin{tabular}{lcccc}
        \toprule
        \textbf{Ansatz} & \textbf{FID} & \textbf{KID} & \textbf{IS} & \textbf{Accuracy (\%)} \\
        \midrule
        HQConv          & \textbf{2.57} & \textbf{$0.0008 \pm 0.0003$} & \textbf{$1.7780 \pm 0.0241$} & 81.74 \\
        FQConv          & 3.10          & $0.0013 \pm 0.0004$          & $1.7547 \pm 0.0228$          & 81.31 \\
        Only-Rotations  & 2.91          & $0.0010 \pm 0.0004$          & $1.7693 \pm 0.0227$          & \textbf{82.03} \\
        \bottomrule
    \end{tabular}
    }
\end{table}

\subsection{Additional studies}\label{par:additional_studies}

\textbf{Only Rotation vs Entangled Circuit}: 
To investigate the effect of entanglement, we perform a detailed per-class evaluation comparing two ansatz variants, i.e. only rotation with no entanglement and the HQConv circuits.

Table~\ref{tab:combined_per_class_metrics}, shows that the use of entanglement led to some improvement on the classification performance. Indeed, the F1-scores increases in challenging classes e.g. River (0.72 \textrightarrow{} 0.75). However, in other cases the no-entanglement version achieved slightly higher precision and recall, suggesting that the benefit of entanglement may be context-dependent.



\begin{table}[!ht]
\centering
\caption{Per-class comparison of precision, recall, and F1-score between only rotation with no entanglement (N) and HQConv (E) circuits for QCU-Net.}
\label{tab:combined_per_class_metrics}
\resizebox{\columnwidth}{!}{
\begin{tabular}{lccccccccc}
    \toprule
    \textbf{Class} & \textbf{Prec (N/E)} & \textbf{Rec (N/E)} & \textbf{F1 (N/E)} & \textbf{Support} \\
    \midrule
    AnnualCrop           & \textbf{0.73} / 0.70 & \textbf{0.91} / 0.89 & \textbf{0.81} / 0.79 & 1027 \\
    Forest               & \textbf{0.89} / 0.87 & \textbf{0.67} / 0.54 & \textbf{0.77} / 0.67 & 1027 \\
    HerbaceousVegetation & 0.71 / \textbf{0.72} & \textbf{0.90} / 0.89 & \textbf{0.80} / \textbf{0.80} & 1027 \\
    Highway              & 0.87 / \textbf{0.88} & 0.79 / \textbf{0.80} & 0.83 / \textbf{0.84} & 1027 \\
    Industrial           & 0.92 / \textbf{0.93} & \textbf{0.96} / 0.93 & \textbf{0.94} / 0.93 & 1027 \\
    Pasture              & 0.81 / \textbf{0.82} & 0.80 / \textbf{0.83} & 0.81 / \textbf{0.82} & 1027 \\
    PermanentCrop        & 0.85 / \textbf{0.90} & \textbf{0.61} / 0.60 & 0.71 / \textbf{0.72} & 1027 \\
    Residential          & \textbf{0.97} / 0.96 & \textbf{0.98} / \textbf{0.98} & \textbf{0.97} / \textbf{0.97} & 1027 \\
    River                & \textbf{0.92} /\textbf{ 0.92} & 0.60 / \textbf{0.63} & 0.72 / \textbf{0.75} & 948 \\
    SeaLake              & \textbf{0.68} / 0.62 & \textbf{1.00} / 0.99 & \textbf{0.81} / 0.76 & 948 \\
\bottomrule
\end{tabular}
}
\end{table}

\textbf{Hyperparameters Tuning:}\label{par:tunig}
Table~\ref{tab:risultati} presents the results of the hyperparameters tuning process. The experiments reveal that the learning rate has the most significant impact on model performance. A high learning rate ($10^{-2}$) lead to unstable training and poor generative quality (a very high FID = $206.82$ and low IS = $1.03 \pm 0.002$). Conversely, reducing the learning rate to $10^{-3}$ gives improvements across all metrics (FID dropping to as low as 7.83 and IS rising to $1.89 \pm 0.0268$).

Further fine-tuning of the $\beta_1$ parameter show that the value 1.95 provides the best results (FID = 7.83), indicating improved stability and convergence behavior. The variation of $\beta_2$ show a smaller but still noticeable impact, with values around 0.990 producing better performance.

Than, while keeping the best lr, $\beta_1$, and $\beta_2$ fixed, we noticed that raising the numerical–stability term from $\varepsilon = 10^{-8}$ to $\varepsilon = 10^{-7}$ reduced the FID from $7.83$ to $7.22$ and KID from $3.9\times10^{-3}$ to $3.4\times10^{-3}$, while reducing it to $10^{-9}$ pushed both metrics in the opposite direction. We therefore selected $\varepsilon = 10^{-7}$.

\begin{table}[!ht]
    \centering
    \caption{Results with hyperparameters tuning for the classical baseline.}\label{tab:risultati}
    \resizebox{\columnwidth}{!}{%
    \begin{tabular}{lccccccc}
    \toprule
    \textbf{lr} & $\boldsymbol{\beta_1}$ & $\boldsymbol{\beta_2}$ & $\boldsymbol{\varepsilon}$ 
      & \textbf{FID} $\downarrow$ & \textbf{KID} $\downarrow$ & \textbf{IS} $\uparrow$ \\
    \midrule
    $10^{-2}$ & 1.90 & 0.990 & $10^{-8}$ 
      & 206.8200 & $0.2288\pm0.0016$ & $1.0269\pm0.0020$ \\
    $10^{-3}$ & 1.90 & 0.990 & $10^{-8}$ 
      & 8.2600   & $0.0046\pm0.0008$ & $1.8718\pm0.0233$ \\
      $10^{-4}$ & 1.90 & 0.990 & $10^{-8}$ 
      & 11.5200  & $0.0057\pm0.0010$ & $1.8138\pm0.0269$ \\
      \midrule
    $10^{-3}$ & 1.85 & 0.990 & $10^{-8}$ & 8.0300   & $0.0040\pm0.0007$ & $1.8977\pm0.0231$ \\
    $10^{-3}$ & 1.95 & 0.990 & $10^{-8}$ & 7.8300 & $0.0039\pm0.0007$ & $1.8922\pm0.0268$ \\
    \midrule
    $10^{-3}$ & 1.95 & 0.895 & $10^{-8}$ & 8.4902   & $0.0048\pm0.0009$ & $1.9005\pm0.0249$ \\
    $10^{-3}$ & 1.95 & 0.995 & $10^{-8}$ & 8.2423   & $0.0042\pm0.0008$ & $1.9033\pm0.0312$ \\
    \midrule
    $10^{-3}$ & 1.95 & 0.990 & $10^{-7}$ & \textbf{7.2248}   & \textbf{0.0034}$\pm$\textbf{0.0008} & $1.8837\pm0.0306$ \\
    $10^{-3}$ & 1.95 & 0.990 & $10^{-9}$ & 8.2325   & $0.0042\pm0.0008$ & \textbf{1.9210}$\pm$\textbf{0.0324} \\
    \bottomrule
    \end{tabular}%
}
\end{table}

\section{Conclusion} \label{sec:conclusion}
In this work, we introduced the integration of quantum computing with class-conditioned diffusion models for generating synthetic EO imagery. By combining the flexibility and modularity of classical architectures, such as U-Net, with quantum circuits, our QCU-Net represents a significant quantum utility for advancement in generative modeling for EO tasks.
Our results highlight that the introduction of quantum layers into the diffusion process considerably enhances both the realism and accuracy of the generated images, especially when these layers are inserted in the encoder branch of the U-Net. Specifically, the QCU-Net model demonstrated marked improvements over both classical and other simpler quantum models, which were included in our analysis to ensure a fair comparison and to highlight the potential utility of quantum framework as proposed. In terms of FID and KID, QCU-Net generates samples that are much closer to the true distribution of real satellite images. Notably, the QCU-Net model showed a 64\% improvement in FID and a 76\% reduction in KID compared to classical models, proving its superior performance. Furthermore, the proposed model exhibited improved semantic accuracy, particularly in challenging classes like AnnualCrop, Industrial and River, that are often difficult for conventional models due to high inter-class variability.
A unique contribution of this study is the class-conditioning mechanism, which enables the generation of labeled EO imagery directly applicable to real-world remote sensing tasks, such as data augmentation or class-specific simulation. This is particularly valuable in scenarios where labeled data are scarce or costly to acquire.
By combining class-conditioned QDMs with U-Net architectures, our approach demonstrates how quantum circuits can significantly enhance generative modeling, especially in remote sensing, where high-dimensional data and labeled images are often scarce.
The results of our ablation studies support this, revealing that the integration of quantum layers in the middle layers of the model, within the feature extraction stages, leads to a notable performance boost. This finding underscores the importance of strategically placing quantum circuits within the model to achieve optimal results. 
Additionally, the HQConv quantum circuit ansatz, which incorporates entangling gates, proved to be effective in capturing the complex dependencies inherent in EO images, further emphasizing the role of quantum entanglement in enhancing generative models. However, in some cases the no-entanglement version achieved slightly better results, suggesting that the benefit of entanglement may be context-dependent and need further investigation.

Beyond the reported quantitative improvements, the proposed QCU-Net model also offers 
tangible practical benefits for real-world EO workflows. First, by enabling class-conditioned 
generation of high-quality synthetic labeled imagery, the model directly mitigates the shortage of annotated EO datasets, which are costly and time-consuming to acquire. This makes the approach particularly valuable for training deep models in low-data regimes or for enriching 
rare land-cover classes.

Second, because the method produces semantically and statistically consistent synthetic samples, it can support large-scale EO data synthesis, improve classification robustness, and 
facilitate downstream tasks such as domain augmentation, rare-class boosting, and simulation of unobserved conditions in operational settings.

To offer future perspectives, we want to highlight some challenges of the proposed approach.

The QCU-Net, while outperforming the classical approach, still relies on classical components for certain stages, especially in managing high-dimensional feature maps in the encoder.
Therefore, it   necessitates careful optimization of both quantum and classical layers, leading to a tradeoff between performance and computational feasibility. Moreover, the reliance on the EuroSAT RGB dataset, though widely used, limits the generalizability of our results to other EO datasets that may differ in spectral bands or spatial resolutions. Additionally, the computational demands of quantum-classical hybrid models require substantial resources, posing a potential barrier to wider adoption, especially given the current limitations of NISQ devices.

Looking ahead, future research should explore additional quantum features, such as quantum noise-driven diffusion and quantum error correction, which could address the limitations of current NISQ devices and enhance model robustness. Furthermore, the scalability and efficiency of QDMs across various quantum hardware platforms remain critical areas of investigation, as the computational demands of quantum circuits are still a significant challenge. Finally, extending this work to multi-modal EO datasets that incorporate other data sources, such as SAR or LiDAR, could facilitate more comprehensive, multi-sensor EO applications.\\
In any case, the proposed framework has demonstrated to represent a powerful tool for the EO domain and community of remote sensing researchers, paving the way  for quantum-enhanced applications in remote sensing.


\section*{Acknowledgments}
The authors gratefully acknowledge the computing resources from the DEEP-EST project, which received funding from the European Union's Horizon 2020 research and innovation programme under the grant agreement no. 754304. The work of F. Mauro and S.L. Ullo was (partially) supported by the European Union under the Italian National Recovery and Resilience Plan (PNRR) of NextGenerationEU partnership on "Telecommunications of the Future" (PE00000001 – program "RESTART"), CUP E63C22002040007 - D.D. n.1549 of 11/10/2022.
The contribution of F. De Falco, A. Ceschini and M. Panella in this work was in part supported by the ``NATIONAL CENTRE FOR HPC, BIG DATA AND QUANTUM COMPUTING'' (CN1, Spoke 10) within the Italian ``Piano Nazionale di Ripresa e Resilienza (PNRR)'', Mission 4 Component 2 Investment 1.4 funded by the European Union - NextGenerationEU - CN00000013 - CUP B83C22002940006.

\bibliographystyle{IEEEtran}
\bibliography{main}

\end{document}